\documentclass[onecolumn]{elsart}
\usepackage{amssymb}
\usepackage{amsfonts}
\usepackage{amsmath}
\usepackage{graphicx}
\usepackage{subfigure}
\usepackage{epsfig}
\usepackage{hyperref}
\usepackage{color}

\input{tcilatex}

\begin{document}

\begin{frontmatter}%

\title{Status of the differential transformation method}%

\author{C.\ Bervillier}\ead{claude.bervillier@lmpt.univ-tours.fr}%

\address{Laboratoire de Math\'{e}matiques et Physique Th\'{e}orique,\\ UMR 7350 (CNRS),\\
F\'ed\'eration Denis Poisson,\\
Universit\'{e} Fran\c{c}ois Rabelais,\\
Parc de Grandmont, 37200 Tours, France}%

\begin{abstract}%
Further to a recent controversy on whether the differential
transformation method (DTM) for solving a differential equation is purely
and solely the traditional Taylor series method, it is emphasized that the
DTM is currently used, often only, as a technique for (analytically) calculating the power series of the
solution (in terms of the initial value parameters). Sometimes, a
piecewise analytic continuation process is implemented either in a
numerical routine (e.g., within a shooting method) or in a semi-analytical
procedure (e.g., to solve a boundary value problem). Emphasized also is the
fact that, at the time of its invention, the currently-used basic
ingredients of the DTM (that transform a differential equation into a
difference equation of same order that is iteratively solvable) were already
known for a long time by the ``traditional''-Taylor-method users (notably in the elaboration
of software packages --numerical routines-- for automatically solving ordinary differential
equations). At now, the defenders of the DTM still ignore
the, though much better developed, studies of the ``traditional''-Taylor-method users who, in turn,
seem to ignore similarly the existence of the DTM. The DTM has been given an apparent strong formalization
(set on the same footing as the Fourier, Laplace or Mellin
transformations). Though often used trivially, it is easily attainable and easily adaptable to different kinds of differentiation
procedures. That has made it very attractive. Hence applications
to various problems of the Taylor method, and more generally of the power
series method (including noninteger powers) has been sketched. It seems
that its potential has not been exploited as it could be. After a discussion
on the reasons of the ``misunderstandings'' which have caused the controversy,
the preceding topics\ are concretely illustrated. It is concluded that, for the sake of clarity, when
the DTM is applied to ODEs, it should be mentioned that the DTM exactly coincides
with the traditional Taylor method, contrary to what is currently written.%
\end{abstract}%


\end{frontmatter}%

\section{Introduction}

The differential transformation method (DTM) of Pukhov \cite%
{7277}--\cite{7260bis} and Zhou \cite{7259} is frequently presented as a
(relatively) new method for solving differential equations\footnote{An extensive presentation
of the DTM is given in section \ref{sec:DTM2}.}. Though based on
Taylor series, it would be different from the traditional Taylor (or power)
series method presented in usual textbooks as e.g., \cite{7309}. This
distinction has been the object of a recent dispute \cite{7272,7273}.
Independently of whether this distinction was present in the original
ideas\footnote{%
The most frequently cited original works on DTM are inaccessible to me
because they are written either in \ Russian \cite{7260,7260bis} or in Chinese \cite%
{7259}. Except those two references, I have systematically solely cited
articles written in English and, hopefully, accessible to all.\label{Foot1}},
it is at least seemingly clearly expressed and often repeated since the
second half of the 90's when the DTM has more systematically been used
\textquotedblleft to solve differential equations\textquotedblright\ \cite%
{7302}--\cite{7375}. For example, in \cite{7267} the 
\textsl{\textquotedblleft differential transformation
technique\textquotedblright }, is presented\textsl{\ \textquotedblleft as an
extended Taylor series method\textquotedblright } and in \cite{7236}, where
the differential equation referred to is 
\begin{equation}
\frac{dx}{dt}=f\left( x,t\right) \,,  \label{eq:equadif0}
\end{equation}%
one can read (in addition):

\textquotedblleft \textsl{The differential transformation technique is one
of the numerical methods for ordinary differential equations. It uses the
form of polynomials as the approximation to exact solutions which are
sufficiently differentiable. This is in contrast to the traditional
high-order Taylor series method, which requires the computation of the
necessary derivatives of }$\QTR{sl}{f(x,t)}$\textsl{\ and is computationally
intensive as the order becomes large. Instead, the differential
transformation technique provides an iterative procedure to get the
high-order Taylor series. Therefore, it can be applied to the case with high
order.}\textquotedblright\ \cite[p. 25]{7236} (see also \cite{7375})

That statement is important since then the idea that the ``traditional'' Taylor series
method would require the explicit calculation of high-order derivatives of $%
f\left( x,t\right) $ is repeated in many articles on the DTM as if it would
be a convincing reason to make a definitive distinction between the two
methods.

Before going any further, it is worth recalling what is the Taylor series method\footnote{For the sake
of a convenient writing, extensive descriptions
of the ``traditional'' Taylor series method and of the DTM are postponed to section \ref{sec:DTM}.}:
 \begin{definition}[(Formal or raw) Taylor series method]
The Taylor series method consists in expressing the solution of (\ref{eq:equadif0}) as a power series
expansion about the initial time $t_0$:
\begin{equation}
\left. 
\begin{array}{l}
x\left( t\right) =\sum_{k=0}^{\infty}\left( t-t_{0}\right) ^{k}\frac{1}{k!}\left. \frac{d^{k}x}{dt^{k}}\right\vert _{t=t_{0}} \,, \\ 
x\left( t_{0}\right) =\alpha _{0}\,.%
\end{array}%
\right\}  \label{eq:Tayl00}
\end{equation}
in which the derivatives $\left. \frac{d^{k}x}{dt^{k}}\right\vert _{t=t_{0}}$ are such that (\ref{eq:equadif0}) is satisfied
order by order in powers of $\left( t-t_{0}\right)$. As consequence, the Taylor coefficients of the expansion are completely determined
once the initial parameter $\alpha _{0}$ is fixed.
\label{rawTaylor}
\end {definition}

Notice that this definition specifies neither the way the derivatives $\left. \frac{d^{k}x}{dt^{k}}\right\vert _{t=t_{0}}$ are calculated nor
the convergent property of the series. Thus
it is surely clear to all that the $\emph{formal}$ Taylor
series method (described in textbooks such as \cite{7309}) requires only that the
series of $x\left( t\right) $ be obtained by any means whatsoever.
Consequently, it is obvious that if the DTM is a \textquotedblleft \textsl{%
procedure to get the high-order Taylor series\textquotedblright } \cite%
{7236,7375}, then the method used to solve the differential equation is not
the DTM but the Taylor series method (at least formally). Consequently to really appreciate the above
quotation of \cite{7236}, it is necessary to understand what actually is the ``traditional'' Taylor series method. No doubt that for the DTM users,
the naive explicit computation of high order
derivatives of $\QTR{sl}{f(x,t)}$ is an integral part of the ``traditional'' method. One of the object of the present article is to show that this allegation is false.

In fact, as one may anticipate from
the presence of the terms \textquotedblleft \textsl{technique%
\textquotedblright } and \textquotedblleft \textsl{numerical
method\textquotedblright }, the Taylor series method mentioned in the above
quotation of \cite{7236}, though not explicitly defined, is not the \emph{%
formal} one\footnote{Which is the best known formulation of the Taylor method.}. It would be possible that it is a problem of semantics which has
generated misunderstanding.

Those words (technique, numerical) have wittingly been used to qualify the
DTM. Indeed, some authors would like to see the DTM not as a formal method,
but exclusively as a new \emph{numerical} approach:

\textsl{\textquotedblleft ... the construction of power-series solutions has
been generally thought of as an analytic tool and not as the basis for
numerical algorithms. This is changing, and algorithmically constructing
power-series solutions to ordinary differential equations is gaining in
popularity. This is now often called the Differential Transform(ation)
Method (DTM).\textquotedblright\ }\cite{7243}

However, if this view corresponds well to Pukhov's initial aim of
\textquotedblleft \textsl{consider}[ing] \textsl{the problem of feasibility
of constructing computer-specialized procedures oriented toward automatic
solution of Taylor equations}\textquotedblright \cite{7277}, the actual use
of power-series solution as the basis for numerical algorithms is not at all
new, see, for example, \cite{7332}--\cite{7336}. Actually:

 \begin{remark}[(Numerical) Taylor series method]
Taylor series may be used as a tool to numerically solve the initial value problem
associated with (\ref{eq:equadif0}). To this end the convergence of the Taylor series (\ref{eq:Tayl00}) must be controlled.
The most frequently used procedure is the stepwise (or piecewise) procedure
described in section \ref{sec:StepWise}.
\label{numTaylor}
\end {remark}

In this respect of the numerical treatment of ODEs, and considering the
high level of numerical development of the traditional Taylor method (see section \ref{HighLevel}), one is forced to acknowledge that
the current use of the DTM cannot be seen as a real technological break-through (despite the huge number of publications on the subject).
One may easily verify that,  the
DTM (as it is currently used) brings no new capability into the numerical
treatment of (at least ordinary) differential equations (see section \ref%
{sec:DTM}). Consequently, to possibly distinguish
it from the traditional Taylor series method only remains the way the expansion coefficients are calculated, in
apparent accordance with the above quotation of \cite{7236}.

Unfortunately, when looking at the
``old papers''  
using the Taylor method, in that respect of calculating the terms of the Taylor expansion, one quickly realizes that the
basical tricks of the DTM (that transform the differential equation into an
algebraic iteration equation, see section \ref{sec:DTM}) were already known
and used (for a long time, indeed, see section \ref{sec:Getting}) without having any need to change the name of the Taylor series
method (e.g., see \cite{7332,7332bis,7341}). Even some efficient procedures of the traditional approach have remained ignored by the DTM users (see section 
\ref{sec:usual}). Several articles that present numerical
softwares using the traditional Taylor series method with a view to
automatically solve systems of ODEs are still currently published \cite%
{7368}--\cite{7340}. Based on an old and well established method
(see, e.g. \cite{7350}), the authors of these articles have no need to
mention the existence of the DTM, though they use exactly the same process!
In fact, the reverse should be the rule: the DTM users should cite traditional studies,
what they never do, probably because they do not know their existence.

The status of the DTM is made very confused because that aspect of the DTM to be potentially\ a \textsl{%
computer-specialized procedure} is often ignored by several DTM users. Most
authors using the DTM do not implement its numerical ability, they provide
the solutions to the problems studied under merely raw Taylor series. When
such studies are directed towards systems of pure ODEs, the original
contribution of the DTM is practically vacuous since it obviously coincides with
 extremely simple, to not say trivial, traditional studies (see section \ref{sec:DTM}).

This explains and justifies the recent biting comment \cite{7272} in which
one may read: \textquotedblleft \textsl{the DTM is the familiar Taylor
series method with a fancy name}\textquotedblright\ and also:

\textquotedblleft \textsl{There is no doubt that the DTM is merely the
textbook Taylor series approach that some authors have disguised as a fancy
discrete transformation.}\textquotedblright\ \cite{7272}

The sharpness of the misunderstanding is well emphasized by the reply \cite%
{7273} to that comment:

\textquotedblleft \textsl{Several revered authors and reputed journals}
[would] \textsl{have nurtured this `misconception' over the years. It}
[would be] \textsl{a challenge to such authors and journals who have
published several works on DTM.}\textquotedblright\ \cite{7273}

It is time to clarify the status of the DTM. This is the tentative object of
the present article.

As one may\ have already foreseen, there are two aspects of the DTM, it may
be seen either

\begin{enumerate}
\item[(A)] as a \textquotedblleft \textsl{computer-specialized procedure}%
\textquotedblright\ (as proposed in \cite{7277} for example),
\end{enumerate}

or

\begin{enumerate}
\item[(B)] as a technique for
determining the terms of the raw Taylor\footnote{%
More generally of power series including non-integer power series.} series
of solutions of various problems (see \cite{7237} as an example among
several others),
\end{enumerate}

or both, of course.

Concerning interpretation A, it is worth recalling that, in general, the
process of solving a differential equation using a Taylor (power) series
requires two steps:

\begin{enumerate}
\item Obtening the solution as a formal power series (Taylor series,
whatever the technique used to get it).

\item The eventual implementation of an analytic continuation process when
the range of convergence of the direct sum of the series is too limited to
reproduce the solution sought in its expected domain of definition. This
procedure may be used:

\begin{enumerate}
\item either, in a numerical routine to solve an initial value problem (or
within a shooting method procedure) with fixed initial parameters

\item or in a semi-analytic approach of, e.g., a boundary value problem (the
series is expressed in terms of the unknown --yet undetermined-- initial
parameters, see e.g., \cite{6889}).
\end{enumerate}
\end{enumerate}

Now, it is true that, from its origin, the\ DTM implicitly involves the
possibility of automatically implementing an analytic continuation via a
piecewise process (although it is said in terms closer to the numerical
version of item 2a, as a multistage numerical method, see for example \cite%
{7277}). This possibility has been exploited by several authors 
(see \cite{7238,7267,7375,7243} and \cite{7266}--\cite{7242}). In that case, the method is truly
different from the raw Taylor series method which, stricto sensu, does not
involve any consideration of analytic continuation (see for example \cite%
{6889}). One may then rightly consider the method as being \textquotedblleft 
\textsl{an extended Taylor series method}\textquotedblright \cite{7267}.
But, as explicitly shown in section \ref{sec:DTM}, in the circumstances, the
DTM merely uses the \textquotedblleft \textsl{analytic continuation}%
\textquotedblright\ process of the Taylor series, as already Euler did it
(see \cite{7488}) and which is described in textbooks under the name of
\textquotedblleft \textsl{continuous analytic continuation}%
\textquotedblright\ \cite[chap. 9]{7179} or of \textquotedblleft \textsl{%
Taylor algorithm}\textquotedblright\ \cite[p. 267]{7452}. This is a
technique included for a long time and still currently used in numerical
softwares for solving ODEs by the ``traditional'' Taylor series method (see, \cite%
{7332}--\cite{7340} and \cite{7445,6557}).

There is a more serious problem with the way that the DTM is viewed by some
authors. This problem is linked with  the interpretation B and
the analytic continuation presently in
question.

Many studies of ODEs using the DTM have been applied according to interpretation B
to so simple problems  that the solutions
sought were analytical (at least in the range of supposed interest of the
independent variable, it is, in particular, the case of the study criticized
in \cite{7272}). In such circumstances, where no analytic continuation was
needed or performed, one has employed the expression DTM (the name of the
technique used to get the raw Taylor series) as if it was a new (formal) method
in place of the Taylor series method itself. This unfortunate view of the
DTM is, by now, so common that when an analytic continuation process of the
series has been realized by means of Pad\'{e} approximants, one has changed
the name of the DTM into the \emph{modified} differential transformation
method (MDTM or also the \textquotedblleft DTM-Pad\'{e} method%
\textquotedblright ) \cite{7280}--\cite{7254}. Such names are not justified because
the DTM enables, in principle\footnote{%
It is possible (see footnote \ref{Foot1}) that this eventuality has been
forseen in the original presentation of the method, but --to my knowledge--
this has never been used by the defenders of the DTM in the current
literature, hence the new name of MDTM.}, the automatic implementation of Pad%
\'{e} approximants (see section \ref{sec:MDTM}). Anyway, in the
circumstances, the (formal) method used is merely the Taylor series
analytically continued by Pad\'{e} approximants (see, for example, \cite%
{6889}) a method which has been used in numerical routines for a long time 
\cite{7353}.

Even more serious is the recent and repeated proposal of
introducing, as a novelty, a piecewise procedure within the DTM and of
naming this \textquotedblleft new method\textquotedblright\ again MDTM (with
\textquotedblleft M\textquotedblright\ for modified or multistage or
multistep) \cite{7317}--\cite{7399}. The authors of such proposals (and the referees
who have allowed their publications) were manifestly unaware of the fact
that the procedure was already clearly foreseen in the original DTM (see,
for example, \cite{7277}) and already implemented by several DTM users \cite%
{7238,7267,7375,7243}, \cite{7266}--\cite{7242}, not to mention the previous
utilizations (some are cited above) by the traditional-Taylor-method users.

Extensions of interpretation B to problems which are potentially more tricky than a simple
system of ODEs are seemingly more interesting (despite the standing formal
character of this interpretation). As a matter of fact, the DTM has been
used as a technique for determining the terms of
the raw power\footnote{%
Including non-integer power series.} series of solutions of various
problems. Its basic ingredients, though not new in the current use, have
been presented in such a manner that it has appeared very attractive (see section \ref{sec:DTM2}). 
Rid of the original constraining goal of performing finite precision calculations, its
implementation, sufficiently easy and clearly presented, has been formally extended to several
different kinds of equations. Thus, equations of the following types have been considered
using the DTM: integro-differential \cite{7311,73111}, partial-differential \cite%
{7258}, \cite{7250}--\cite{72509}, differential-difference \cite{7254,7314,73141},
differential-algebraic \cite{7310,73101}, fractional order differential \cite%
{7399}, \cite{7355}--\cite{735515}, fuzzy differential \cite{7383}, q-partial differential \cite%
{7400}, and so on. Of course, some of those subjects had already been
similarly (i.e. analytically) addressed by Taylor-method users, but the lack of a systematic formalization of the
traditional approach could have limited their impact (see \cite{7516}--\cite{7451}, for purely analytical approaches).
However, if one considers studies with finite precision goal then, again, the traditional method has reached
a higher level of development such as in the treatment of the differential
algebraic case --a particular system of ODEs already well treated \cite%
{7368,7350}, \cite{7359}--\cite{73593}-- or of integral and integro-differential equations (see, e.g., \cite{75163,7422}).
I will not discuss\ further the use of the DTM in such problems.
Nevertheless, it seems that the DTM could deserve its name (as a technique) when it extends
the Taylor method to new kinds of expansions (different from a Taylor
expansion, see e.g., \cite{7356}) and to unusual derivation processes such as
in the q-calculus. Nevertheless, in such studies, considering their formal character limited to the obtaining the algebraic iteration equations for the coefficients
of the series expansion, it should be clearly stated that the
\emph{method} used (to solve the primary equation) remains
merely the Taylor (or power) series method.

Moreover, with interpretation B, the series is calculated only once
(and most often it is limited to low orders).
Consequently the recourse to an iterative procedure to determine the raw power series is even not
necessary. Thus, in this meaning, the recourse to DTM may appear entirely artificial (a ``fancy'' name). As already said,
that version of the DTM is so commonly used
that it is by now often presented as ``the'' DTM.
That interpretation, surely, has induced the polemic and the confusions mentioned
above. This does however not give more credit to interpretation A, as shown in what follows.

The following section is limited to the problem of
solving ODEs (numerically or not).  It is explicitly shown that the traditional Taylor
method\ should take precedence over the DTM.  It is also indicated why it is not
acceptable to modify the name \textquotedblleft DTM\textquotedblright\ when
the resulting Taylor series is analytically continued.

\section{The Taylor series method vs. the differential transform method \label%
{sec:DTM}}

In this section, a comparison is made between the most elaborated
current formulations of the Taylor series method and of the DTM when they
are applied to solve an initial (or boundary) value problem of ODEs (for
recent reviews on the use of the Taylor method see, e.g., \cite%
{7365}--\cite{7340} and \cite{7361} and for the most elaborated presentations of the
DTM see, e.g., \cite{7245,7238,7267,7242,7374}). Only the one dimensional case
of\ the first order ODE (\ref{eq:equadif0}) is considered here,
the generalization to several variables and to systems of ODEs of higher
orders is straightforward.

\subsection{The Taylor series method\label{sec:TSM}}

The recourse to ``infinite series'' to solve differential equations is an old method
which, as mentioned in \cite[p. 4]{7488},
may be attributed to Newton in \cite{7580}. Since then, the method
has been rediscovered several times (as well illustrated, notably, in \cite{7339}). By now, it is named the Taylor series method.
Several current papers may be found that present this method in details
(for example: \cite{7350,7340,7361}, see also \cite[pp. 47--49]{7488}). The
following considerations are thus far from being original.

For the sake of considering practical solutions of (\ref{eq:equadif0}), and with a view to describe the ``traditional'' Taylor method, definition
\ref{rawTaylor} given p. \pageref{rawTaylor} must be slightly modified.

With the Taylor series method, the generic solution of (\ref{eq:equadif0}) is
expressed locally as a \emph{truncated} Taylor expansion about a time $t_{i}$:%
\begin{equation}
\left. 
\begin{array}{l}
x\left( t\right) =\sum_{k=0}^{N}\left( t-t_{i}\right) ^{k}X_{k}+O\left[
\left( \left( t-t_{i}\right) ^{N+1}\right) \right] \,, \\ 
x\left( t_{i}\right) =\alpha _{i}\,, \\ 
X_{k}=\frac{1}{k!}\left. \frac{d^{k}x}{dt^{k}}\right\vert _{t=t_{i}},\quad
k=1,2,\cdots ,N\,.%
\end{array}%
\right\}  \label{eq:Tayl0}
\end{equation}

In the following it is assumed that $t>t_{i}$, and $\alpha _{i}$ is supposed
to be known either because $t_{i}$ is the initial time (say $t_{0}$) at
which the initial condition is provided or because it is an intermediate
time value ($t_{i}=t_{0}+i\,h$) reached by a stepwise analytic continuation
procedure (see section \ref{sec:StepWise}). Although not indicated here and in
the following (except where noted), the coefficients $X_{k}$ depend on $%
t_{i} $. The dependence is implicit via the initial value $\alpha _{i}$, but
it may also be explicit if the ODE considered is non-autonomous.

As said in the introduction, the method involves two steps: first
the obtaining the series for some (high?\footnote{%
Actually, it is rarely necessary to consider high values of $N$, as is shown
in the following.}) value of $N$, and second a process of analytic
continuation of the series. Stricto sensu, the second step is not a part of
the formal Taylor series method. But, because it is by now difficult to
distinguish between studies based on Taylor series which are formal (getting
only the raw series) or numerical (using, in addition, analytic continuations),
I temporarily include the two steps in the \emph{traditional} Taylor series
method (awaiting to show below why that is justified), reserving the name 
\emph{raw} or \emph{formal} Taylor series method exclusively to the first
step (in accordance with definition \ref{rawTaylor} p. \pageref{rawTaylor}).

\subsubsection{Getting the raw Taylor series\label{sec:Getting}}

The coefficients $X_{k}$ must be determined such that the ODE (\ref%
{eq:equadif0}) be formally satisfied order by order up to $O\left[ \left(
\left( t-t_{i}\right) ^{N}\right) \right] $, one condition\footnote{%
In general the order is $N-p$, with $p$ the order of the ODE, and $p$
conditions are required to unambiguously determine the $X_{k}$.} ($x\left(
t_{i}\right) =\alpha _{i}$) is required so that the resulting system of $N$
equations for the $X_{k}$ ($k=0,1,\cdots ,N$) has a definite solution.

Establishing and solving that system of equations is not the most efficient
way of determining the coefficients $X_{k}$ (in particular this would imply
the explicit calculation of \textquotedblleft \textsl{the necessary
derivatives of }$\QTR{sl}{f(x,t)}$\textquotedblright ). In fact, it is known
for a long time by the (thus traditionnal) Taylor-series-method users that
the $X_{k}$ may be determined more efficiently by iteration provided $%
f\left( x,t\right) $ be simple enough (see, e.g. \cite{7393,7367,73671}, see also 
\cite[pp. 47--49]{7488}). Instead of $N$ equations, only one equation (the
iteration equation) has, then, to be effectively considered.

Assuming, as a simple illustration, that $f\left( x,t\right) =\lambda
\,x\left( t\right) $ with $\lambda $ a constant, it is easy to show that the
generic Taylor coefficients $X_{k}$ satisfy the following iteration equation:%
\begin{equation}
\left. 
\begin{array}{l}
X_{0}=\alpha _{i}\,, \\ 
\left( k+1\right) \,X_{k+1}=\lambda \,X_{k}\,,\quad k=0,1,\cdots ,\infty
\,\,.%
\end{array}%
\right\}  \label{eq:DiffEq1}
\end{equation}

This is a difference equation of\ same order (first) as the ODE under study.
With the initial condition, the coefficients $X_{k}$ are obtained easily by
iteration of (\ref{eq:DiffEq1}) up to high orders, with e.g.:%
\begin{equation*}
X_{0}=\alpha _{i}\,,\quad \,X_{1}=\alpha _{i}\lambda \,,\quad \,X_{2}=\alpha
_{i}\frac{\lambda ^{2}}{2}\,,\quad \cdots \,,\quad X_{N}=\alpha _{i}\frac{%
\lambda ^{N}}{N!}\,,\quad \cdots \,.
\end{equation*}

The\ terms of the Taylor series of an exponential is recognized giving the
exact solution (which satisfies the right initial condition):%
\begin{equation*}
x\left( t\right) =\alpha _{i}\,\mathrm{e}^{\lambda \left( t-t_{i}\right) }\,.
\end{equation*}%
Although that example is trivial, it indicates the way towards a more
elaborated procedure when $f\left( x,t\right) $ is more complicated. The
above example may be presented as follows. One constructs a priori a table
of correspondence:%
\begin{equation*}
\begin{array}{ccl}
x\left( t\right) & \rightarrow & X_{k}\,, \\ 
\frac{dx\left( t\right) }{dt} & \rightarrow & \left( k+1\right) \,X_{k+1}\,,
\\ 
\lambda \,x\left( t\right) & \rightarrow & \lambda \,X_{k}\,,%
\end{array}%
\end{equation*}%
so that the ODE of interest is automatically transformed into a difference
equation:%
\begin{equation*}
\begin{array}{ccc}
\frac{dx\left( t\right) }{dt}=\lambda \,x\left( t\right) & \rightarrow & 
\left( k+1\right) \,X_{k+1}=\lambda \,X_{k}\,.%
\end{array}%
\end{equation*}

Generalizing to the $k^{\mathrm{th}}$ Taylor coefficient of $f\left(
x,t\right) $ (denoted below by $F_{k}$) then it comes:%
\begin{equation}
\left( k+1\right) \,X_{k+1}=F_{k}\,.  \label{eq:Fk}
\end{equation}

To construct the $F_{k}$'s, the Taylor-series-method users have considered
supplementary obvious (or already known) correspondences such as (see \cite[%
p. 525]{7384} but also \cite{7387}):%
\begin{equation}
\begin{array}{ccl}
\frac{d^{n}x\left( t\right) }{dt^{n}} & \rightarrow & \left( k+n\right)
\cdots \,\left( k+1\right) X_{k+n}\,, \\ 
x\left( t\right) +y\left( t\right) & \rightarrow & X_{k}+Y_{k}\,, \\ 
x\left( t\right) \,y\left( t\right) & \rightarrow & \sum_{i=0}^{k}\,X_{i}%
\,Y_{k-i}\,, \\ 
z\left( t\right) =\frac{x\left( t\right) }{y\left( t\right) } & \rightarrow
& \frac{1}{Y_{0}}\left( X_{k}-\sum_{i=0}^{k-1}Z_{i}Y_{k-i}\right) \,, \\ 
z\left( t\right) =\left[ x\left( t\right) \right] ^{\beta } & \rightarrow & 
\sum_{i=1}^{k}\left( \frac{\beta +1}{k}i-1\right) \frac{X_{i}}{X_{0}}%
Z_{k-i},\quad k\geq 1,\quad Z_{0}=X_{0}^{\beta }\,.%
\end{array}
\label{eq:Table1}
\end{equation}

One has then considered more complicated forms of $f\left( x,t\right) $ such
as exponential (similar to $\left[ x\left( t\right) \right] ^{\beta }$),
logarithm, trigonometric functions etc...\cite{7332,7332bis} (see also \cite[pp. 48,
49]{7488}) so as to find similar simple formulas for their images. The aim
was, notably, to include them into numerical softwares \cite{7405} with a view to solve
automatically various kinds of systems of ODEs \cite{7332,7332bis,7406}.

As noted in\ \cite{7336},\ such recurrence schemes for the terms of a Taylor
series in solving a differential equation was used as early as 1946 by
J.C.P. Miller \cite{7393}. Some authors (see, e.g. \cite{7408}) attribute
the first use of such recurrences to the computation of derivatives by J.R.
Airy in 1932. Actually,  the iterative procedure was known already at the time of Newton
(see, \cite[Prob. II, Sol. Case II, Ex.I, p. 33]{7580} and also, e.g., \cite[p. 116]{7579}), and  the last line of (\ref{eq:Table1}), often attributed
to J.C.P. Miller \cite[p. 507 of second ed.]{7384} and \cite{7385}, had been established
by L. Euler in 1748 \cite{7388}, cf \cite[p. 526]{7384}. Such a formula is
the result of the application of the Leibnitz rule for the derivative of a
product of functions. This rule, and others, is at the basis
of an efficient procedure for
numerically calculating high order derivatives named automatic
differentiation (for extensive bibliographies, see \cite{7583,75831}),
which is used in turn to get efficient iterative formulas for complicated forms of $f(x,t)$.
This technique is used for a while in numerical routines \cite%
{7368}, \cite{7362}--\cite{73626}, \cite{7340,7408}, \cite{7421}--\cite{7421bis1} and in regular studies (e.g., \cite{6557}). More
recently, similar considerations have led to a proposal \cite{7523} for
efficiently implementing the Picard iteration method.

It is worth indicating that such recurrences (and\ others) have been
established also for functions of two variables (see e.g., \cite{7334}  for analytical expressions) and that the automatic
differentiation process in many variables is also well developed for numerical applications (see, e.g., \cite{7671,7608,76081}). 

In the following,\ the name "Taylor transformation" will designate the above
described automatic procedure for obtaining the iteration equation (the
image) of a given ODE (by the way, it was the primary name of the DTM \cite%
{7277} a name rightly re-used in the first papers of the 90's \cite%
{7302,7245,7238}).

It is important to keep in mind that, up to now, the above rules of calculation
of the terms of the Taylor series is purely formal. In particular it does not anticipate the nature
 of computation (theoretical exact or finite precision) which is to be done. In the
following subsections, the considerations are less formal and are progressively more and more oriented towards numerical applications.

\subsubsection{Analytic continuations}

Sometimes the generic Taylor series is convergent in the domain of
definition of interest (such as in the preceding trivial case of the
exponential). The (formal) method then provides an exact solution (when $N=\infty $
and if the generic term is identified) or, at least, a convergent form
towards the exact solution (for $N$ finite). Those simple cases, often
linked to linear ODEs, are not of great interest.

More interesting are the nonlinear ODEs,\ the generic Taylor series of the
solutions of which have finite radius of convergence due to the presence of
singularities in the complex plane of the independent variable $t$. The
efficiency of the Taylor expansion thus is limited to a finite domain\
(which may be small) about the expansion point $t_{i}$. To enlarge this
domain, the recourse to analytic continuation procedures is often required.
(Notice however that in practice, one may need to use analytic continuation even
for series with an infinite radius of convergence, if only to reduce the length of the series.)

There are several kinds of analytic continuations. Let us consider only the
two most often used in conjunction with the Taylor series method.

\paragraph{Stepwise procedure \label{sec:StepWise}}

The stepwise procedure, also named the \textquotedblleft \textsl{continuous
analytic continuation}\textquotedblright\ \cite[chap. 9]{7179} or the
\textquotedblleft \textsl{Taylor algorithm}\textquotedblright\ \cite[p. 267]%
{7452}, is particularly interesting because, beyond its character of
analytic continuation procedure, it is structurally well adapted to
numerical routines.

Let the domain of definition of the solution sought, say $t\in \lbrack
t_{0},t_{0}+H],$ be cut in $n\in 
\mathbb{N}
$ pieces: $H=n\,h$. Then the times of reference read $t_{i}=t_{0}+i\,h$, $%
i=0,1,\ldots ,n$. If $h$ is chosen sufficiently small (at each step, smaller
than the distance to the closest complex singularity of the solution), then
the Taylor series converges in the domain $\left[ t_{i},t_{i+1}\right] $, and
its sum provides the solution in this domain with a sensible accuracy even
if $N$ is not very large. If this is true for each value $t_{i}$ then one
gets, step by step, an approximate solution in the whole domain $[0,H]$,
each initial value $\alpha _{i}$ being (approximately) provided by the sum
of the Taylor series at the second boundary of the preceding sub-domain,
namely

\begin{equation}
\alpha _{i}=\sum_{k=0}^{N}h^{k}\left( X_{k}\right) _{\left( i-1\right)
}+O\left( h^{N+1}\right) \,,  \label{eq:alphai}
\end{equation}%
in which I have indicated the dependence of $X_{k}$ on the step $i-1$.

This stepwise procedure has been (and is currently) successfully used in
numerical routines by Taylor-series-method users to solve initial value
problems (for reviews see, e.g. \cite{7365,7350,7340}). Its first use in
conjunction with power series goes back to Euler \footnote{%
Euler's motivation was not to perform an analytic continuation but,
precisely in opposition to the statement of \cite{7236} quoted in the
introduction, to avoid the calculation of high-order derivatives of $f\left(
x,t\right) $.} \cite{7488}. For that reason, it is justified to consider\ it
as being an integral part of the traditional Taylor series method as the
name \textquotedblleft \textsl{Taylor algorithm}\textquotedblright\ \cite[p.
267]{7452} suggests it.

The procedure has also been used to solve boundary value problems \cite%
{7375,7368,7406,7447}. In that case, the unknown $\alpha _{0}$ of the
initial boundary is transmitted, step by step via the parameters $\alpha
_{i} $, up to the second boundary $t_{0}+H$ where the desired condition is
imposed. The value of the unknown is chosen among the acceptable solutions
of the resulting equation. The method is limited by two aspects:

\begin{enumerate}
\item The dependence of $\alpha _{i}$ on the unknown becomes rapidly very
complicated. (However, see \cite{7587,7634} in which supplementary Taylor expansions in powers
of the initial unknown are used as generally done when using the Taylor method in a sensitivity analysis, see e.g., \cite{7608,76081}.)

\item When $H\rightarrow \infty $ the method no longer works because, when $%
t\rightarrow \infty $ the truncated Taylor series goes to $\pm \infty $
according to the sign of the last term
\end{enumerate}

In the last case, where the domain of definition of the problem is infinite,
it is better to consider other kinds of analytic continuation such as, for
example, rational fractions (Pad\'{e} approximants).

\paragraph{Pad\'{e} approximants}

The use of Pad\'{e} approximants is well known \cite{7110,71101,71102}, and will not\ be
described here. It suffices to recall that it is a procedure that transforms
a truncated series (a polynomial) into a ratio of polynomials. It is a so
drastic modification of the Taylor series of the generic solution that it
cannot be easily seen as the \textquotedblleft inverse\textquotedblright\ of
an \textquotedblleft image\textquotedblright\ resulting from some
\textquotedblleft modified\textquotedblright\ Taylor transformation.
However, it is shown in section \ref{sec:MDTM} that, similarly to the
step-wise procedure, the recourse to Pad\'{e}-approximant seems to have been
also foreseen in the DTM but, seemingly, has never been implemented in the
current literature.

\paragraph{High level numerical development\label{HighLevel}}
There is no doubt that, as a numerical tool,
 the traditional Taylor series method has reached a high level of developement. I have already mentioned several numerical routines that have been developed with a view
to automatically solve the initial (and boundary) value problems of systems of ODEs \cite%
{7368}--\cite{7340}. I must mention also the existence of adjustments of the Taylor method 
to treat stiff systems \cite{7368},\cite{7362}--\cite{7350}, \cite{7482}--\cite{74824} and the possibility that this method offers to get high-precision solution for ODE \cite{7631}. In the same vein, it is worth mentioning the use of Taylor 
series methods in interval (or validated) solutions, that provide a garantee of existence of (and bounds on) the solution
in a given interval \cite{7421bis,7421bis1}, \cite{7593}--\cite{75934} (for a bibliography on the subject of validated solutions see, e.g., the paper by Nedialkov et al in \cite{7421bis,7421bis1}). 
As an extension to those latter studies, it is fair to mention also the developement of new methods for obtaining garanteed bounds on the expansion
\cite{7587}, \cite{7659}--\cite{76591}, for a review see Makino and Berz in \cite{7645}.

In this respect of the quality of the numerical treatment of ODEs the difference is glaring with studies using the DTM
in which almost never are raised questions like, e.g., selecting the
stepsize (or the order of development) or discussing the computational complexity of the system considered.
In particular, despite the original Pukhov's aim, I have not seen any paper that proposes a numerical routine with a view to automatically solve
systems of ODEs
nor a possible adaptation of the DTM to treat stiff systems.

\subsection{The differential transform method\label{sec:DTM2}}
In this section the most elaborated version\footnote{Accessible to me, see footnote \ref{Foot1}.} of the DTM
 and its usual version are presented and compared with the Taylor series method
presented in the preceding sections.

\subsubsection{General procedure\label{sec:general}}

The DTM \cite{7277}--\cite{7259} is a formalized modified version of
the Taylor transformation described in section \ref{sec:TSM}. The basic
difference is that instead of $X_{k}$ defined in (\ref{eq:Tayl0}), one
considers the image $\breve{X}_{k}$ of $x\left( t\right) $ defined as \cite%
{7245,7238,7267,7242,7374}: 
\begin{equation}
\breve{X}_{k}=M_{k}\left. \frac{d^{k}\left( q\left( t\right) x\left(
t\right) \right) }{dt^{k}}\right\vert _{t=t_{i}}\,,  \label{eq:DTM000}
\end{equation}%
in which the auxiliary function $q\left( t\right) $ and the infinite set $%
M_{k}$ are chosen a priori. $M_{k}$ is called the weighting factor and $%
q\left( t\right) $ is named the kernel corresponding to $x\left( t\right) $
(see, e.g. \cite{7374}).

One recovers $x\left( t\right) $ through the following series (the inverse
transform):%
\begin{equation}
x\left( t\right) =\frac{1}{q\left( t\right) }\sum_{k=0}^{\infty }\frac{%
\left( t-t_{i}\right) ^{k}}{k!}\frac{\breve{X}_{k}}{M_{k}}\,.  \label{eq:xq}
\end{equation}

\subsubsection{Usual procedure\label{sec:usual}}

To my knowledge (see footnote \ref{Foot1}), the freedom in the choice of $%
q\left( t\right) $ has never been exploited in the current literature, where
it is usually set equal to one (for possible uses of $q\left( t\right) $,
see section \ref{sec:MDTMPade}). Moreover, the choice of $M_{k}$ is
currently limited exclusively to the two following cases (except in \cite%
{7242} where an adaptive grid size mechanism is employed):

\begin{itemize}
\item $M_{k}=\frac{1}{k!}$, then $\breve{X}_{k}$ $\equiv X_{k}$ (since $%
q\left( t\right) \equiv 1$), the DTM coincides with the Taylor
transformation and the method (to solve a differential equation) to the
formal Taylor method (without any analytic continuation --i.e., the interpretation B mentioned in the introduction).

\item $M_{k}=\frac{h^{k}}{k!}$ as\ proposed in \cite{7277,7252}, where $h$
has the same meaning as in section \ref{sec:StepWise} and is effectively
used in the same way (see, e.g. \cite{7252}). One has $\breve{X}%
_{k}=h^{k}\,X_{k}$, and eq. (\ref{eq:alphai}) is slightly
simplified:%
\begin{equation}
\alpha _{i}=\sum_{k=0}^{N}\left( \breve{X}_{k}\right) _{\left( i-1\right)
}+O\left( h^{N+1}\right) \,.  \label{eq:alphaiDTM}
\end{equation}%
The drawback is that one must account for the presence of $h$ in the
definitions of the images such as, e.g. 
\begin{equation*}
\begin{array}{ccl}
x\left( t\right) & \rightarrow & \breve{X}_{k}\,, \\ 
\frac{dx\left( t\right) }{dt} & \rightarrow & \frac{\left( k+1\right) }{h}\,%
\breve{X}_{k+1}\,.%
\end{array}%
\end{equation*}%
It is easy to verify that one obtains the same result by merely rescaling
the time variable $t\rightarrow h\,t$ within the traditional Taylor approach.
Although, in practice, there may be advantages one way
or the other\footnote{Indeed, this form
has been included in some routines of Taylor-series-method users to reduce underflow and overflow
in practical computation
by choosing $h$ of the order of the radius of convergence of the series \cite{7336}.}, this is too little to justify a drastic change of designation of
the method used. 
\end{itemize}

For the remaining aspects of the DTM, that is to say the way that $f\left(
x,t\right) $ is treated, there is no difference from the Taylor
transformation described in section \ref{sec:Getting}. Only simple forms of $%
f\left( x,t\right) $ are considered, and each case is treated separately.
Finally a table of correspondence similar to (\ref{eq:Table1}) is
constructed. 

It is important to mention that, concerning the construction of the table of correspondence, DTM users have recently been reminded of
the existence of the exponentiation
recurrence [last line of (\ref{eq:Table1})] with a view to improve the
efficiency of their method \cite{7318}.\ The same kind of remark had already
been made some time before in \cite{7500,75001} where the recourse to the Leibnitz
rule of derivation is used to get the image of the exponential or logarithm
functions through \textquotedblleft \textsl{new algorithms}%
\textquotedblright \footnote{%
Already known by the users of the traditional-Taylor-series method.} which
are more efficient than those of the standard DTM. These proposals for
improving the DTM clearly contradict the claim quoted in the introduction
that the traditional Taylor series method would be \textsl{\textquotedblleft
computationally intensive as the order becomes large\textquotedblright }
compared to the DTM since, for a long time, those recurrences were known (see, e.g. \cite{7387}
and section \ref{sec:Getting}) and already
included in routines to solve ODEs by Taylor series (see, e.g. \cite%
{7334,7336}).
These late propositions for improving the DTM, and the succes of the DTM itself, show enough that, as the traditional Taylor method,
 the automatic differentiation procedure \cite{7583}--\cite{7421bis1} is largely unrecognized
in the current litterature, this is a pity.

\subsubsection{The modified differential transform methods \label{sec:MDTM}}

\paragraph{Modification by Pad\'{e} approximants\label{sec:MDTMPade}}

When the domain of definition of the solution sought is infinite ($H=\infty $%
), some authors have proposed substituting for the stepwise procedure
foreseen in the DTM by using Pad\'{e} approximants \cite{7280}--\cite{7254}. To this
end one first chooses $M_{k}=\frac{1}{k!}$, whereas $q\left( t\right) $ is
still set equal to one (i.e. the raw Taylor series is obtained first). One
then transforms the Taylor series (the inverse image) into a rational
fraction (according to the rule of Pad\'{e} calculus \cite{7110}--\cite{71102}). Now,
because the recourse to Pad\'{e} approximants (eventually after a Laplace
transform of the series \cite{7403,74031}) has been implemented at the level of
the inverse of the image of $x\left( t\right) $, then this is no longer a
differential transformation (modified or not).

In fact, eq. (\ref{eq:xq}) shows that the DTM enables the automatic
implementation of the Pad\'{e}-approximant technique. Indeed, it suffices to
decide that $q\left( t\right) $ is a polynomial in $\left( t-t_{i}\right) $
of order $N_{2}$ (with coefficients $Q_{k}$, $k=1,\cdots ,N_{2}$ to be
determined and $Q_{0}=1$) and to limit the $\breve{X}_{k}$-series to be of
order $N_{1}$ with $N_{1}+N_{2}=N$, to get a balanced system of equations
for the $Q_{k}$ and $\breve{X}_{k}$ expressed directly in terms of the
\textquotedblleft image\textquotedblright\ of $f\left( x,t\right) $ and of
the initial value $\alpha _{i}$. To be concrete, the system of equations for
the $Q_{k}$'s and $\breve{X}_{k}$'s is: 
\begin{equation}
\begin{array}{ll}
\sum_{i=0}^{N_{2}}Q_{i}\,X_{k-i}=0,\quad & k=N_{1}+1,\cdots ,N_{1}+N_{2}\,, \\ 
\breve{X}_{k}=\sum_{i=0}^{N_{2}}Q_{i}\,X_{k-i},\quad & k=0,\cdots ,N_{1}\,,%
\end{array}
\label{eq:Q}
\end{equation}%
in which $Q_{0}=1$ and the $X_{k}$'s are implicitly known from (\ref{eq:Fk})
in terms of the initial value $\alpha _{i}$.

Notice that, because the first line of (\ref{eq:Q}) is a system of $N_{2}$
linear equations which cannot be solved iteratively, it is necessary to
determine the $X_{k}$'s explicitly. Consequently, the process is not
more efficient than applying the Pad\'{e}-technique on the Taylor series
and, in any way, the method used remains merely the Taylor series
analytically continued by Pad\'{e} approximants.

An other possible use\footnote{%
Notice also that $M_{k}$ could be used to implement a Borel transform of the
Taylor series of $x\left( t\right) .$} of $q\left( t\right) $, in the case
of an infinite value of the range $H$, could be as follows. Sometimes one
knows that, when $t\rightarrow \infty $, the solution sought behaves as $%
t^{\nu }$ with $\nu \in 
\mathbb{Q}
$, and this behavior is hardly reproduced by a Taylor series even continued
by a Pad\'{e} approximant. To circumvent this difficulty, one may
choose (assuming $t_{i}=0$):%
\begin{equation}
q\left( t\right) =\left( 1+\frac{t}{r}\right) ^{\emph{-}\nu }  \label{eq:qnu}
\end{equation}%
in which $r$ is a parameter to be adjusted in order that the radius of
convergence $\left\vert r\right\vert $ of the series expansion of (\ref%
{eq:qnu}) be larger than that (a priori unknown) of the Taylor series of $%
x\left( t\right) $. With such a choice for $q\left( t\right) $, the behavior
at infinity of the transformed function is a simple constant reproducible by
a diagonal Pad\'{e} approximant.

\paragraph{Modification by the stepwise procedure}

I have already expressed, in the introduction, what must be thought of the
proposal of improving the DTM by introducing a stepwise analytic
continuation of the Taylor series, when this procedure was already clearly
foreseen in the DTM and used several times as recalled in section \ref%
{sec:usual}.

\subsubsection{Formalization}

With a bit of emphasis, the interpretation B of the DTM has been 
presented on the same footing as the Laplace, Fourier, and Mellin transforms
where the integration procedure used to construct the image of a given
function is replaced by a derivation procedure \cite{7252} (see also \cite%
{7249}). The inverse transformation gives a power series (when $q\left(
t\right) =1$ and $M_{k}=\frac{1}{k!}$, it is merely the Taylor series of the
function). By similitude with the Laplace etc. transformations, a list of
properties of the DTM with respect to the addition, multiplication, etc. has
been given, as well as the images of current functions. 
The resulting transformation of, say, an ODE into an algebraic equation easily solvable, has been
presented as being formally equivalent to that of the Laplace (or Fourier) transform
 with the additional advantage that the DTM applies also to nonlinear ODEs.

Of course, the depth of the considerations is lesser than in the traditional Taylor series approach. But this latter has
 been almost
exclusively
developed as a numerical tool. This ability of the Taylor series method being not very well publicized, 
the formal presentation of the DTM
has appeared (wrongly) as a new and elegant way of constructing a table of correspondence like (\ref{eq:Table1}).

As emphasized enough in the present article, given the high level
of developement of the traditional Taylor series method (as a numerical tool), the DTM
contributes nothing new in the way systems of ODEs are solved. Nevertheless,
the DTM has allowed an easy generalization of the
Taylor method to various derivation procedures. For example, fractional
differential equations have been considered using the DTM extended to the
fractional derivative procedure \cite{7399}, \cite{7355}--\cite{735515} via a modified version of
the Taylor series \cite{7356}. In similar situations --such are also fuzzy
differential \cite{7383}, q-differential \cite{7400}, etc.-- the DTM might
deserve its name (although traditional Taylor series treatments also exist
that deal with fractional derivative \cite{7451} or fuzzy equations 
\cite{7517,75171}). 

Less evident is the original contribution of the DTM to the study of partial
differential, integro-differential, integral, etc. equations where the
usual derivative procedure is in action (e.g., integral and
integro-differential equations have been studied by Taylor series methods in 
\cite{7516}--\cite{73461}, \cite{7422}). 

In general, it would be fair to not forget the limit of the formal version of the DTM which is not
a method to solve the initial problem but only a convenient way of obtaining the iteration equation
for the power series coefficients (provided the rules of the automatic derivation be well accounted for).
For the sake of clarity, in such situations, a clear
reference to the traditional Taylor (or power) method should be mandatory.

\section{Conclusion}

A misunderstanding on what the \textquotedblleft
traditional\textquotedblright\ Taylor series method is has grown out since
the 90's of a desire to promote an attractive method named the differential
transformation method. Actually, when the DTM was being born, the Taylor
series method had developed for quite a while the treatment of ODEs exactly
on the same basis and, concerning this treatment, absolutely nothing new has
originated from the new method. Even some efficient recurrence formulas
(known for a while) have been unknown to the DTM users.

The misunderstanding has increased over time because the DTM
(and also the Taylor series method, of course) has two possible uses: as an
analytic tool or as a numerical tool. Although it was created with this latter
use in mind, the DTM has most often been used as an analytic tool, so that
some\ authors have even forgotten the initial aim and proposed to reinvent
the method.

Beyond the treatment of ODEs with the DTM --in which case one should at
least clearly refer to the numerous studies done with the Taylor series
method-- it seems that the major contribution of the DTM is in the easy
generalization of the Taylor method (either analytical or\ numerical) to
problems involving unusual derivative procedures such as fractional, fuzzy
or q-derivative.

\textbf{Acknowledgments } I thank Prof.\ Saeid Abbasbandy for several useful
remarks on this work and the reviewers who have indicated me several ways for
improving this article. I thank also Bruno Boisseau, Hector Giacomini and Stam Nicolis
for their kind encouragement and support.

\end{document}